\documentclass[journal]{IEEEtran}
\usepackage{amsmath,amsfonts}
\usepackage{algorithmic}
\usepackage{algorithm}
\usepackage{array}
\usepackage[caption=false,font=normalsize,labelfont=sf,textfont=sf]{subfig}
\usepackage{textcomp}
\usepackage{stfloats}
\usepackage{url}
\usepackage{verbatim}
\usepackage{graphicx}
\usepackage{cite}
\usepackage{balance}
\usepackage{array}
\usepackage{pifont}
\usepackage[table]{xcolor}
\usepackage[hidelinks]{hyperref}
\usepackage[T1]{fontenc}
\begin{document}

\title{Towards Quantum SAGINs Harnessing Optical RISs: Applications, Advances, and the Road Ahead}

\author{Phuc V. Trinh,~\IEEEmembership{Senior Member,~IEEE,} Shinya Sugiura,~\IEEEmembership{Senior Member,~IEEE,} 
Chao~Xu,~\IEEEmembership{Senior Member,~IEEE,} and Lajos~Hanzo,~\IEEEmembership{Life Fellow,~IEEE}
\thanks{Preprint (Accepted Version). DOI: 10.1109/MNET.2025.3536848.
Copyright $\copyright$ 2025 IEEE. Personal use of this material is permitted. However, permission to use this material for any other purposes must be obtained from the IEEE by sending a request to pubs-permissions@ieee.org.}
\thanks{P. V. Trinh and S.~Sugiura are with the Institute of Industrial Science, The University of Tokyo, Tokyo 153-8505, Japan.}
\thanks{C. Xu and L. Hanzo are with the School of Electronics and Computer Science, University of Southampton, SO17 1BJ Southampton, U.K.}
\thanks{This work was supported in part by Japan Science and Technology Agency (JST) Adopting Sustainable Partnerships for Innovative Research Ecosystem (ASPIRE) under Grant JPMJAP2345; in part by Japan Society for Promotion of Science (JSPS) KAKENHI under Grant 23H00470, Grant 24K17272, and Grant 24K21615; and in part by the Telecommunications Advancement Foundation (TAF).}
\thanks{L. Hanzo would like to acknowledge the financial support of the Engineering and Physical Sciences Research Council (EPSRC) projects under grant EP/Y037243/1, EP/W016605/1, EP/X01228X/1, EP/Y026721/1, EP/W032635/1, EP/Y037243/1 and EP/X04047X/1 as well as of the European Research Council's Advanced Fellow Grant QuantCom (Grant No. 789028).}
}

\markboth{Preprint (Accepted Version) for publication in IEEE Network (DOI: 10.1109/MNET.2025.3536848)}%
{Trinh \MakeLowercase{\textit{et al.}}: Towards NG Quantum SAGINs with Optical RIS}


\maketitle

\begin{abstract}
The space-air-ground integrated network (SAGIN) concept is vital for the development of seamless next-generation (NG) wireless coverage, integrating satellites, unmanned aerial vehicles, and manned aircraft along with the terrestrial infrastructure to provide resilient ubiquitous communications. By incorporating quantum communications using optical wireless signals, SAGIN is expected to support a synergistic global quantum Internet alongside classical networks. However, long-distance optical beam propagation requires line-of-sight (LoS) connections in the face of beam broadening and LoS blockages. To overcome blockages among SAGIN nodes, we propose deploying optical reconfigurable intelligent surfaces (ORISs) on building rooftops. They can also adaptively control optical beam diameters for reducing losses. This article first introduces the applications of ORISs in SAGINs, then examines their advances in quantum communications for typical SAGIN scenarios. Finally, the road ahead towards the practical realization of ORIS-aided NG quantum SAGINs is outlined.

\end{abstract}


\section{Introduction}
\IEEEPARstart{N}{ext}-generation (NG) networks are anticipated to provide abundant resources to meet the escalating traffic demands of immersive services. However, relying solely on terrestrial communication systems cannot support flawless wireless access in complex urban scenarios or in challenging environments such as oceans and mountains. Fortunately, the emergence of the space-air-ground integrated network (SAGIN) concept holds the promise of seamless global connectivity in support of advanced applications \cite{Wang2024}. By incorporating satellites, unmanned aerial vehicles (UAVs) \cite{Xu2019}, manned aircraft, and the terrestrial infrastructure, SAGIN ensures resilient and ubiquitous communications even in remote areas of economic activity. This integration enhances network capacity, reliability, and mobility, supporting the Internet-of-Things, autonomous vehicles and telemedicine. Additionally, SAGINs facilitate secure long-range communications via quantum key distribution (QKD) \cite{Chen2024}, paving the way for a synergistic quantum Internet (Qinternet) that operates alongside classical networks \cite{Trinh2024}. 

In the Qinternet, photonic quantum bits, which utilize photons to carry quantum information, are well-suited to both optical fiber and free-space optical (FSO) networks \cite{Li2021}. FSO systems transmit data via laser beams through the ether for establishing direct line-of-sight (LoS) connections without fiber. Using FSO to deliver quantum photons within SAGINs would support the ``Qinternet in the sky", which involves low-Earth orbit (LEO) satellites, high-altitude platforms (HAPs), and low-altitude platforms (LAPs) \cite{Trinh2024}. Entanglement resources are utilized for the transfer of quantum information across these non-terrestrial platforms and distribution to terrestrial networks. Initially, entanglement links are dynamically established both within and between layers of LEO satellites, HAPs, and LAPs, and then extended to ground segments. User requests may trigger the creation of end-to-end entanglement through entanglement swapping, connecting any desired nodes within the quantum SAGINs. End users on the ground or non-terrestrial platforms may then exploit this entanglement for quantum communications \cite{Li2023}. Once the end-to-end entanglement is established, quantum information is directly transferred between end nodes, bypassing intermediate nodes, regardless of the distance between them.

In quantum SAGINs, HAPs typically involve UAVs in the stratosphere at around 20 kilometers (km), while LAPs consist of fixed-wing drones, helicopters, and rotary-wing drones operating at various altitudes below 5 km, depending on national flight regulations. Additionally, manned airplanes flying at an average altitude of 9 km between HAPs and LAPs can also be integrated into SAGIN, creating a complex three-dimensional network of interconnected non-terrestrial vehicles. While LEO satellites may transmit directionally towards platforms at diverse altitudes, HAPs and LAPs carry their payloads underneath using gimbals. As the upper part of HAPs is reserved for solar panels and that of LAPs houses critical components like batteries and global positioning system (GPS) antennas, placing communication terminals on top would reduce the solar panel area on HAPs and interfere with GPS signals and experience severe vibrations on LAPs. Additionally, top-mounted weights increase the risk of tipping, especially in strong winds. By contrast, mounting equipment underneath HAPs and LAPs reduces vibrations via gimbals and facilitates LoS links with lower-altitude targets. Nevertheless, this practical configuration hinders the transmission and reception of optical signals to and from higher-altitude platforms.

To address this practical challenge, we propose deploying optical reconfigurable intelligent surfaces (ORISs) on building rooftops to enhance FSO communications among various SAGIN platforms. ORISs are passive reconfigurable surfaces that use mirror arrays or metasurfaces to control the orientation and phase shifts of incident optical beams \cite{Ndjiongue2021}. Due to their flat surfaces and compact electronics, ORISs can be conveniently installed on building rooftops that offer high LoS availability to SAGIN platforms. Communication terminals installed beneath HAPs and LAPs can transmit signals towards ORISs, which then reflect the beams upward to other SAGIN platforms. The reflected beams can be adaptively adjusted both in direction and beam diameter to optimize the geometrical and misalignment loss (GML) at the receivers. In this article, we first introduce the applications of ORISs in NG SAGIN for both classical and quantum communications. Then we demonstrate their effectiveness in enhancing the quantum functionalities of SAGINs. Finally, we highlight the road ahead towards the practical realization of NG quantum SAGINs.
\section{ORIS Applications versus Traditional Relays in NG SAGINs}
\subsection{ORIS Applications in NG SAGINs}
\begin{figure*}[t]
\centering
\includegraphics[scale=0.4]{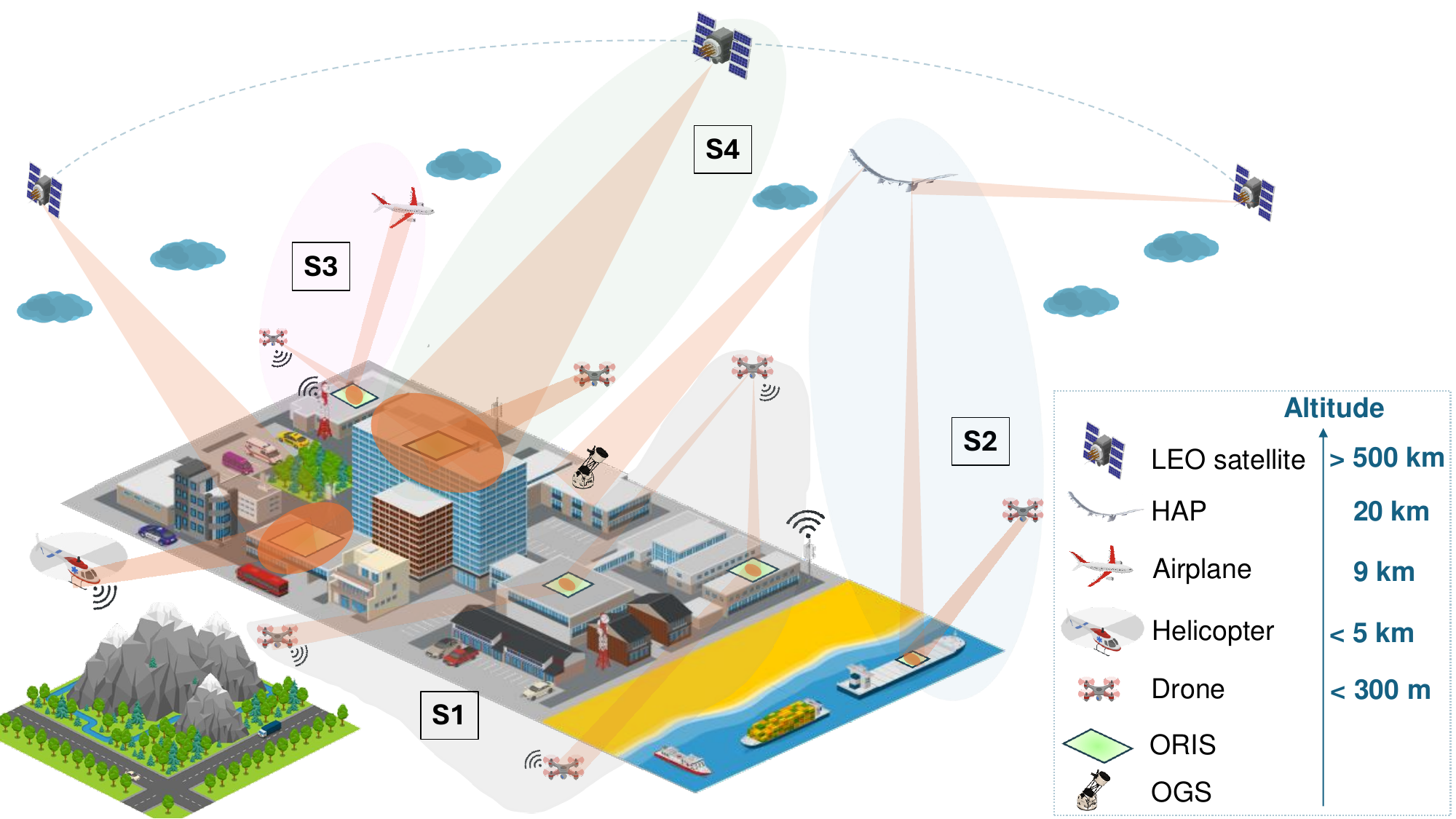}
\caption{Illustration of NG SAGINs with rooftop-based ORISs. Scenario 1 (S1): drone-ORIS-drone links; Scenario 2 (S2): HAP-ORIS-drone links; Scenario 3 (S3): Drone-ORIS-airplane links; Scenario 4 (S4): LEO satellite-ORIS-drone links. }
\label{fig_1}
\end{figure*}
In recent studies, ORISs have been proposed as a solution for both classical \cite{Ajam2024} and quantum \cite{Kundu2024} FSO systems in terrestrial environments, particularly when the direct LoS link is obstructed. ORISs provide a cost-effective alternative to dedicated optical relay nodes by employing passive elements for manipulating the phase of incident beams. This capability facilitates both adaptive beam diameter and angular control at minimal power consumption. The flat surfaces and compact electronics of ORISs facilitate their installation on building walls or rooftops. They typically use either mirror-array or metasurface technologies \cite{Ndjiongue2021}. Mirror-array-based ORISs leverage micro-electro-mechanical systems (MEMS) to adjust the orientation of small mirrors, enabling precise control of optical beams. Nevertheless, MEMS-based ORISs are hindered by narrow beam deflection and low spatial resolution, limiting their effectiveness for large-scale, cost-efficient applications \cite{Trinh2024-JSAC}. By contrast, metasurface-based ORISs utilize smart materials having optically tunable properties, such as liquid crystals, to induce phase shifts by modulating molecular alignments. This enables high spatial resolution, wide beam deflection, and supports scalable designs with continuous tunability \cite{Trinh2024-JSAC}. In addition to enabling the generalized Snell's law of reflection, ORISs efficiently manage beam divergence and focus, serving as pivotal components for connecting platforms in three-dimensional space and optimizing GML at receivers.

In contrast to terrestrial networks, where ORISs are usually mounted on building walls to create near-horizontal links, we propose positioning ORISs on building rooftops. This strategic placement allows for the establishment of non-LoS links among SAGIN platforms by first creating a LoS link with the ORIS, which then reflects towards another SAGIN platform through an additional LoS link. This setup allows for the optimal installation of communication terminals beneath SAGIN platforms, while enabling three-dimensional beam steering in upward directions, as shown in Fig. \ref{fig_1}. Several typical SAGIN scenarios (i.e., S1$\sim$S4) are highlighted for further detailed analyses in Section \ref{sect:adv_ORIS}. Installing ORISs on rooftops eliminates obstructions caused by surrounding structures, ensuring a clearer LoS between ORISs and SAGIN platforms. Additionally, the expansive area available on rooftops provides greater flexibility in positioning the ORISs, facilitating convenient beam-steering and improved connectivity with aerial and satellite platforms. 

\begin{figure*}[t]
\centering
\includegraphics[scale=0.3]{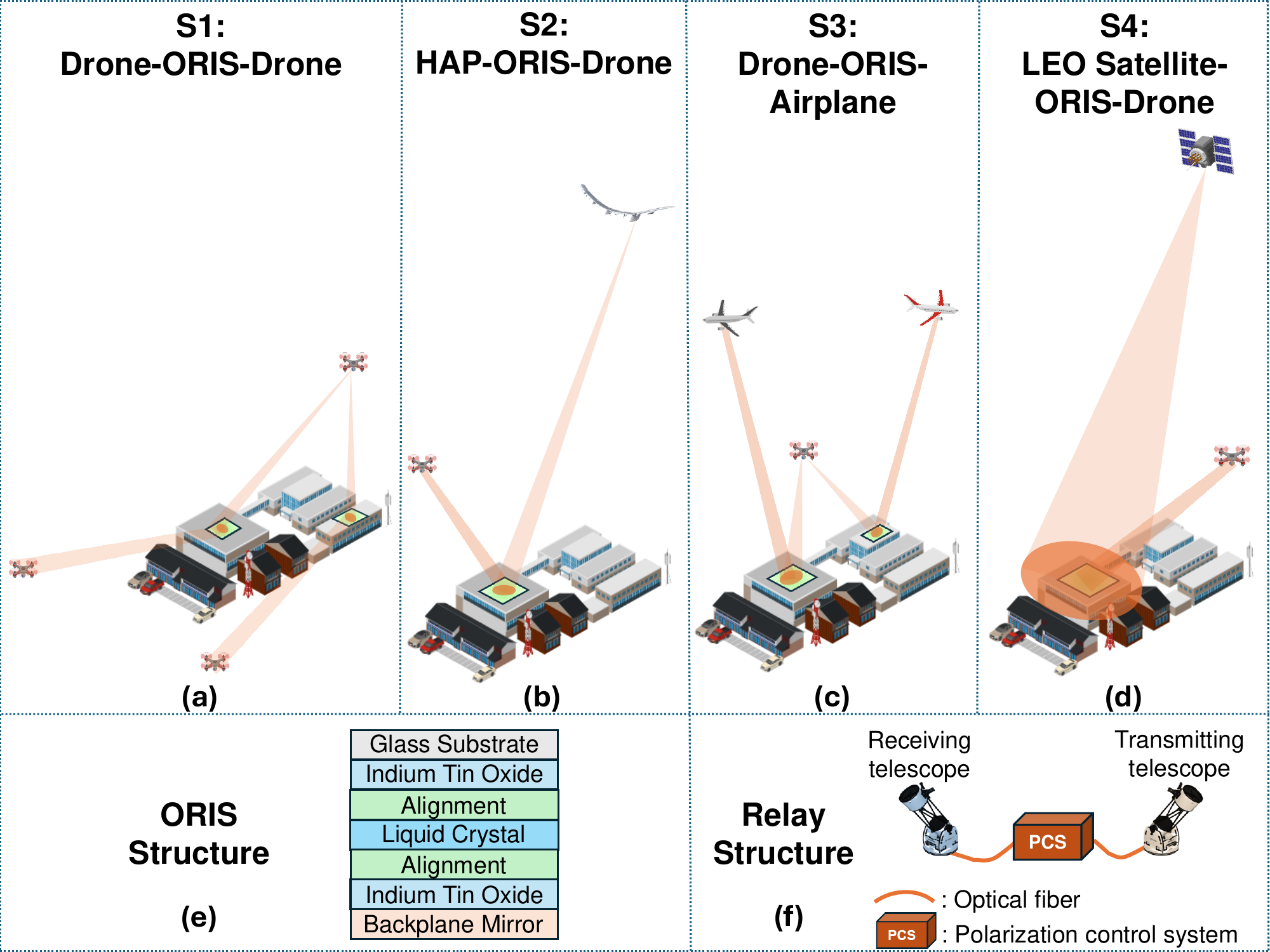}
\caption{Typical scenarios of ORIS-aided NG quantum SAGINs: (a) Quantum entanglement over drone-ORIS-drone links; (b) QKD over HAP-ORIS-drone links; (c) Entanglement distribution over drone-ORIS-airplane quantum links; (d) LEO satellite-ORIS-drone quantum links; (e) Structure of a metasurface-based ORIS; (f) Structure of a traditional relay.}
\label{fig_2}
\end{figure*}
In quantum SAGINs, the space segment of LEO satellites enables global quantum communications with extensive coverage. By circumventing excessive relaying, LEO satellites can efficiently establish LoS links with other satellites at various altitudes or with ORISs. Although establishing LoS links with aerial platforms is challenging due to the downward-oriented nature of terminals beneath these platforms, it remains feasible when satellites are at low elevation angles near the horizon. Once quantum links are established, satellites can act as either trusted or untrusted relays for sharing cryptographic keys via QKD protocols between distant HAPs, LAPs, or ground stations, thereby enabling secure applications \cite{Trinh2024}. Multiple satellites can also serve as quantum repeaters, using entanglement-swapping to extend the quantum communication range. This process may require quantum memories for temporarily storing entangled quantum information \cite{Azuma2023}. Additionally, some satellites may function as quantum routers and switches, facilitating entanglement routing and switching without copying information. 

The air segment includes HAPs, airplanes, and LAPs (e.g., battery-powered drones), which can significantly enhance the quantum network coverage attained in challenging terrains and provide prompt on-demand responses. HAPs, powered by solar panels, are designed for long-duration operations, making them eminently suitable for maintaining continuous communication links. On the other hand, LAPs are ideal for rapid deployment in disaster zones and urban environments, although their operations are typically shorter due to their limited battery life. Notably, the successful demonstration of entanglement distribution for multiple mobile nodes -- with drones acting as an entanglement distribution source and an optical relay -- highlights the growing potential of drones in quantum SAGIN applications [12]. In addition, airplanes serve as a versatile intermediary, effectively bridging the gap between HAPs and LAPs, enhancing their collective capabilities across various applications. HAPs, airplanes, and LAPs are inherently unstable platforms, often hovering or cruising unpredictably around their average positions or trajectories. This instability makes the tasks of pointing, acquisition, and tracking of FSO links particularly challenging. To address this, ORISs act as stable relay points and fixed targets for establishing initial LoS links with HAPs or LAPs. ORISs can then adaptively optimize beam diameters and direct the beams towards desired destinations, including other HAPs, airplanes, LAPs, and satellites.

Finally, the ground segment comprises optical ground stations (OGSs) that function as quantum transceivers or relays. These stations are equipped with telescope apertures flexibly ranging from 0.1 to 1.5 meters (m) and are typically installed on building rooftops or designated areas. The large OGS apertures are crucial for minimizing losses in LoS links from space and air segments to the ground. When connected to optical fiber networks, these OGSs seamlessly distribute received signals into the core networks. On the other hand, ORISs represent an innovative enhancement to ground infrastructure and serve as an alternative to traditional relays, strategically deployed on building rooftops in urban areas to enable non-LoS links among space and aerial platforms. In oceanic and mountainous environments, ORISs can be installed on flat surfaces of ships or trucks. To mitigate the effects of ship motions and truck vibrations, ORISs assisted by stabilization platforms may be required. Fig. \ref{fig_2} summarizes four representative ORIS-aided scenarios in NG quantum SAGINs, including drone-ORIS-drone links (Fig. \ref{fig_2}a), HAP-ORIS-drone links (Fig. \ref{fig_2}b), drone-ORIS-airplane links (Fig. \ref{fig_2}c), LEO satellite-ORIS-drone links (Fig. \ref{fig_2}d). Furthermore, the structures of a metasurface-based ORIS \cite{Ndjiongue2021} and a traditional relay \cite{Liu2021} are depicted in Figs. \ref{fig_2}e and \ref{fig_2}f, respectively, to facilitate comparison in Section \ref{sect:ORISvsRelay}.
\subsection{ORISs versus Traditional Relays}
\label{sect:ORISvsRelay}
Conventionally, to relay quantum signals between non-terrestrial platforms, two adjacent OGSs are interconnected via optical fiber, as depicted in Fig. \ref{fig_2}f. One OGS receives the quantum signals, which are coupled into the optical fiber and linked to the second OGS. At the second OGS, quantum signals, i.e., polarized photons, are corrected for polarization changes introduced by the fiber using a polarization control system composed of multiple wave plates \cite{Liu2021}. The optical beam emitted from the optical fiber is collimated using an optical telescope and then re-transmitted via free-space propagation to the destination. Importantly, no measurement or amplification is performed at the quantum relay, in adherence to the no-cloning theorem.

\begin{table}[t]
\centering
\captionsetup{font=footnotesize}
\caption{Comparison of ORIS-Aided and Relay-Aided SAGINs.}
\scalebox{0.88}{%
\begin{tabular}{l|c|c|}
\cline{2-3}
                      & \textbf{ORIS-Aided SAGINs} & \textbf{Relay-Aided SAGINs} \\ \hline\hline
\multicolumn{1}{|l|}{\textbf{Deployment cost}} & Low  & High  \\ \hline
\multicolumn{1}{|l|}{\textbf{Energy efficiency}} & High  & Low \\ \hline
\multicolumn{1}{|l|}{\begin{tabular}[t]{@{}l@{}}\textbf{Adaptive beam control}\\ \textbf{for optimizing GML}\end{tabular}} & Yes & No \\ \hline
\multicolumn{1}{|l|}{\textbf{Internal loss}} & Low  & High  \\ \hline
\multicolumn{1}{|l|}{\begin{tabular}[t]{@{}l@{}}\textbf{Polarization reference-frame}\\ \textbf{misalignments \& phase delay}\end{tabular}}  & Yes  & Yes  \\ \hline
\multicolumn{1}{|l|}{\textbf{Beam delay spreads}} & Yes  & No  \\ \hline
\end{tabular}%
}
\label{Table_1} 
\end{table}
Table \ref{Table_1} provides a qualitative comparison of ORISs and traditional relays in assisting NG quantum SAGINs. ORISs utilizing liquid-crystal metasurfaces offer the potential for lower costs compared to quantum relays, primarily due to their compact design, reduced number of components, and simplified assembly. By contrast, quantum relays require two large telescopes with complex optical designs, significantly increasing both deployment and operational expenses. Furthermore, ORISs are highly energy-efficient passive devices, relying on a nearby cellular base station to compute phase-shift profiles that direct beams to their targets \cite{Trinh2024-JSAC}. Meanwhile, quantum relays require two telescopes with independent electronics and optical systems to track and align with flying platforms, possibly resulting in higher power consumption. Additionally, quantum relays are designed with fixed beam diameters \cite{Liu2021}, which can cause high GML under severe aerial platform vibrations or hovering fluctuations. ORISs effectively mitigate this issue by dynamically adjusting phase-shift profiles to vary the beam diameters, optimizing GML for changing conditions \cite{Ajam2024,Kundu2024,Trinh2024-JSAC}. Another advantage of ORISs over quantum relays is their lower internal loss, achieved through the use of thin layers of substrates, liquid crystals, and backplane mirrors, as illustrated in Fig. \ref{fig_2}e \cite{Ndjiongue2021}. Conversely, quantum relays suffer considerably higher losses due to fiber coupling inefficiencies and losses introduced by multiple optical components \cite{Liu2021}.

Nevertheless, both ORISs and quantum relays face challenges with polarization reference-frame misalignments and phase delays. These arise from varying ORIS reflection angles \cite{Trinh2024-JSAC} or telescope movements \cite{Trinh2024} in quantum relays, causing misalignments between transmitted and received photon polarizations or turning linear polarization into elliptical. However, these effects may be mitigated by installing a motorized half-wave plate and a quarter-wave plate at the quantum receiver to realign the polarization and eliminate phase differences, respectively \cite{Trinh2024,Trinh2024-JSAC}. In addition, ORIS reflections introduce beam delay spreads, as different reflected components traverse slightly varying distances, resulting in different delays. The delay spread is defined as the difference between the maximum and minimum delay values across all ORIS reflecting elements. Delay spreads increase with ORIS size and the difference between incident and reflected angles \cite{Trinh2024-JSAC}. This may lead to inter-symbol interference if the symbol duration is shorter than the delay spread, requiring careful consideration when utilizing high-speed quantum sources. By contrast, quantum relays with separate telescopes avoid this issue, as the telescopes are aligned and generally perpendicular to the beam's optical axis, ensuring all beam components arrive simultaneously at the receiver.
\section{Advances in ORIS-aided NG Quantum SAGINs}
\label{sect:adv_ORIS}
In this section, we explore representative ORIS-aided quantum SAGIN scenarios depicted in Fig. \ref{fig_2}. 
\subsection{ORIS-Aided Drone-to-Drone Entanglement Links}
\begin{figure}[t]
\centering
\includegraphics[scale=0.55]{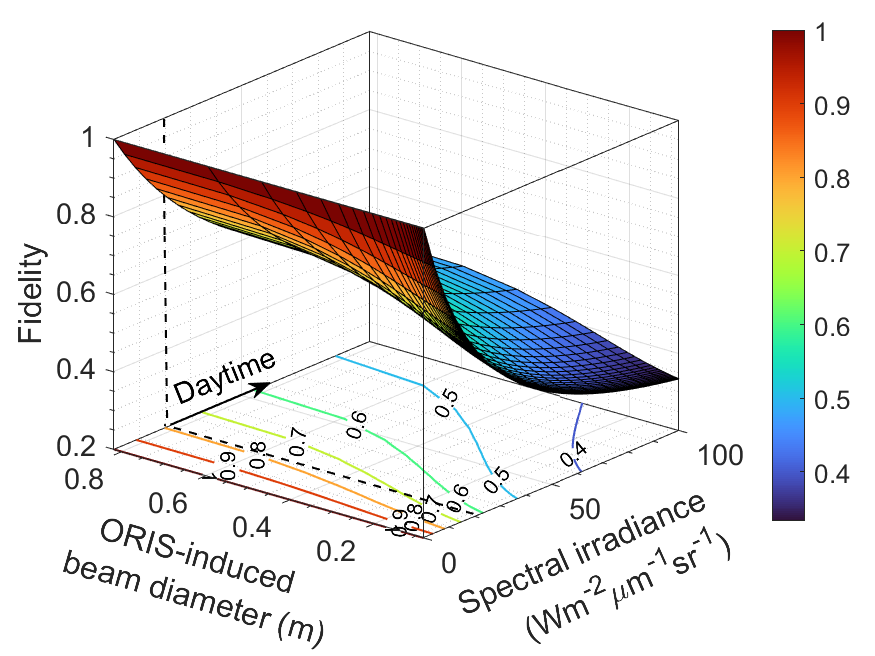}
\caption{Fidelity of ORIS-aided drone-to-drone entanglement transmission at a wavelength of $810$ nanometers (nm) in the presence of drone hovering fluctuations. The transmitted optical beam divergence half-angle is $500$ microradians ($\mu$rad). All terminal apertures have a diameter of $9$ cm.}
\label{fig_3}
\end{figure}
We consider S1, where a drone generates a pair of entangled photons and simultaneously transmits each photon to two separate drones, as illustrated in Fig. \ref{fig_2}a. Although a direct drone-to-drone quantum link has been demonstrated \cite{Liu2021}, it is limited to communications at similar altitudes due to carrying the terminals underneath the drones. Additionally, beam diffraction-induced broadening restricts the communication range due to high GML over long distances. These issues can be effectively addressed by first transmitting the entangled photons to two ORISs located on different building rooftops between the transmitting and receiving drones. Due to these shorter distances, ORISs tend to receive less broadened beams and optimize the reflected beam diameters for the target drones at any arbitrary altitude. This minimizes GML and achieves higher fidelity, an important indicator of the similarity between two shared quantum states \cite{Khatri2021}. 

Following \cite{Trinh2024-JSAC} with the GML framework presented in Section III and channel parameters specified in Table III, we investigate the quantum fidelity of entangled photons using \cite[(33)]{Khatri2021} for the scenario of Fig. \ref{fig_2}a. The transmitting drone is at an altitude of $300$ m, and the receiving drones are at $200$ m. The ORISs are on rooftops at $20$ m. The transmitting and receiving apertures form an elevation angle of $20^{\circ}$ relative to the ORISs, resulting in a total communication distance of about $1.35$ km. We assume that drone hovering fluctuations result in weak pointing errors (PEs) with statistical parameters detailed in \cite[Table III]{Trinh2024-JSAC}. Fig. \ref{fig_3} shows the quantum fidelity results as a function of the ORIS-induced beam diameter and spectral irradiance of background noise. It is observed that the beam diameter can be adaptively adjusted with respect to the spectral irradiance to maximize entanglement fidelity. More specifically, fidelity can be maintained between 80\% and 40\% during daytime.
\subsection{ORIS-Aided HAP-to-Drone QKD Links}
We now investigate S2, where a HAP encodes cryptographic keys into quantum states and transmits them to a drone, as shown in Fig. \ref{fig_2}b. QKD ensures that any eavesdropping attempts to intercept and replicate the transmitted quantum states will inevitably perturb them, revealing the presence of the eavesdroppers. Consequently, the secret key is shared between the HAP and the drone, enabling the safe encryption and decryption of confidential messages. Given that the communication terminals are mounted beneath the HAP and the drone, establishing a direct link is infeasible. If the HAP is positioned far away near the horizon, a near-horizontal direct link with the drone might be possible. However, this poses significant challenges, as the optical beam experiences severe power fluctuations due to atmospheric turbulence and substantial beam broadening. An effective solution in this scenario involves deploying an ORIS on a rooftop. The HAP first establishes a LoS link with the ORIS, which then reflects the incident beam towards the drone. 

\begin{figure}[t]
\centering
\includegraphics[scale=0.55]{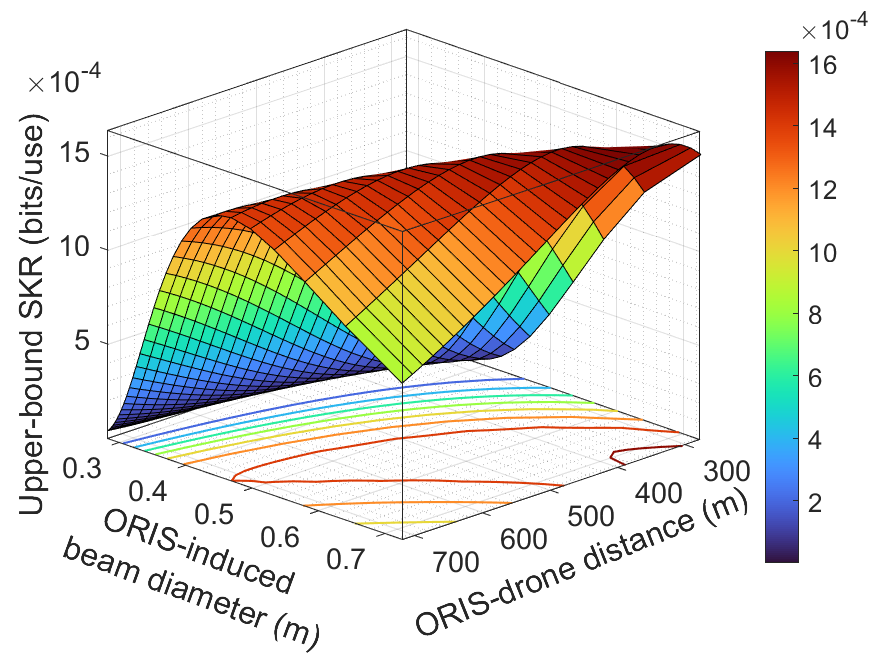}
\caption{Upper-bound SKR of ORIS-aided HAP-to-drone QKD at a wavelength of $810$ nm in the presence of drone hovering fluctuations. The transmitted optical beam divergence half-angle is $16.5$ $\mu$rad. All terminal apertures have a diameter of $9$ cm.}
\label{fig_4}
\end{figure}
In Fig. \ref{fig_4}, we examine the ultimate information-theoretic upper limit for the secret-key rate (SKR) \cite[(19)]{Pirandola2021} of the QKD scenario depicted in Fig. \ref{fig_2}b, considering atmospheric turbulence and weak PE parameters in \cite[Table III]{Trinh2024-JSAC}. The HAP is positioned at an altitude of $20$ km and establishes a direct link with the ORIS at a $25^{\circ}$ elevation angle. The ORIS, deployed on a rooftop $5$ m high, then reflects the incoming beam at a $25^{\circ}$ elevation angle to a hovering drone. The results in Fig. \ref{fig_4} show how the SKR varies with the distance between the ORIS and the drone, when adjusting the ORIS-induced beam diameter. It is evident that selecting an optimal ORIS-induced beam diameter can achieve the desired SKR level for each ORIS-drone distance. Interestingly, we found that the SKR is maximized when the ORIS-drone distance is about $339$ m, with the optimal beam diameter being $73$ cm. This is plausible, since the beam has already traversed nearly 47 km between the HAP and the ORIS. Thus, maintaining a relatively short distance between the ORIS and the drone strikes a balance between minimizing additional losses and ensuring reliability amid drone hovering fluctuations. 
\subsection{ORIS-Aided Drone-to-Airplane Entanglement Links}
We now examine S3, where a drone distributes entangled photon pairs with two airplanes via ORISs, as shown in Fig. \ref{fig_2}c. A direct LoS uplink from the drone to the airplane is impractical due to the terminal's placement beneath the drone. Instead, the drone sends entangled photons to two ORISs on nearby rooftops. These ORISs redirect the photons to the corresponding airplanes. Since the coordinates of the two ORISs are fixed, alignments between the drone and the ORISs become achievable, while the airplanes can be effectively tracked by their respective ORISs via adjusting the phase-shift profiles to direct the optical beams towards the airplanes. Additionally, the typical distances between the ORISs and the airplanes, often spanning dozens of kilometers, may result in significant beam broadening due to diffraction, leading to substantial losses. These losses can be mitigated by the ORISs, which can adjust the beam diameter to counteract the diffraction-induced broadening.

\begin{figure}[t]
\centering
\includegraphics[scale=0.55]{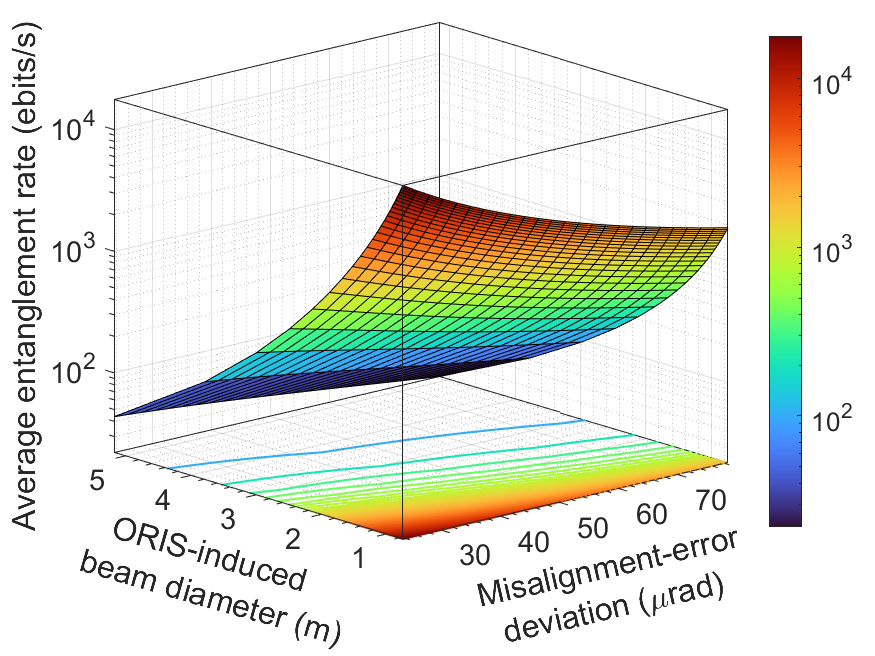}
\caption{Average entanglement rate of ORIS-aided drone-to-airplane entanglement distribution at a wavelength of 810 nm in the presence of link misalignments. The transmitted optical beam divergence half-angle is $500$ $\mu$rad. All terminal apertures have a diameter of 9 cm.}
\label{fig_5}
\vspace{-0.2cm}
\end{figure}

Figure \ref{fig_5} illustrates the average rate of entanglement bits (ebits) received by the airplanes for S3, assuming the drone transmits entangled photon pairs at a rate of $10^9$ ebits/s \cite{Khatri2021}. The drone hovers at an altitude of $300$ m, sending two optical beams at a $45^{\circ}$ angle towards the two ORISs. The airplanes are assumed to fly at an altitude of 9 km, moving in opposite directions at a speed of $720$ km/h (i.e., $200$ m/s). As the airplanes move, point-ahead angles of $1.33$ $\mu$rad must be applied from the ORISs to ensure that the reflected beams can reach the airplanes, assumed at a $45^{\circ}$ angle. The channel models that account for random losses are adapted from \cite{Trinh2024-JSAC}. Additionally, random PEs due to imperfect alignments between the ORISs and airplanes are considered, modeled as Gaussian random variables \cite{Trinh2024-JSAC}. These Gaussian variables are assumed to have zero mean and the same variance for both the horizontal and vertical axes. Various levels of PE variance are explored by examining the misalignment-error deviation in Fig. \ref{fig_5}. Finally, using \cite[(40)]{Khatri2021} and the framework outlined in \cite{Trinh2024-JSAC}, the average ebits are numerically computed. As shown in Fig. \ref{fig_5}, the ORISs effectively reduce the beam diameters reaching the airplanes, thereby improving the entanglement rates under various PE conditions.

\subsection{ORIS-Aided LEO Satellite-to-Drone Quantum Links}
In scenario S4, illustrated in Fig. \ref{fig_2}d, we consider a quantum communication task where a LEO satellite connects with a drone. A direct LoS link to the drone's terminal is impractical because the terminal is positioned underneath the drone. Consequently, the satellite initially directs the optical beam to a strategically positioned ORIS on a building rooftop, which is close to the drone's hovering location. This scenario is the most challenging in quantum SAGINs, as the beam footprint broadens to several hundred meters after traveling hundreds of kilometers from the satellite to the drone. This substantial broadening, compared to the drone's centimeter-scale aperture and its hovering nature, results in a very high GML, thus impeding quantum communications. 

To address this challenge, ORISs offer an innovative solution. A large ORIS can capture a significant portion of the broadened beam and narrow it down to the size of the drone's aperture, thereby significantly improving the GML. The ORIS can be feasibly manufactured to sizes comparable to typical OGS apertures. This capability allows the ORIS to capture the same amount of power as an OGS, enhancing the efficiency and feasibility of satellite-to-drone quantum communications. However, manufacturing large-size ORISs faces significant challenges, particularly in synchronizing the induced phase shifts across the entire surface to maintain consistent optical performance, which requires precise control over nanofabrication processes and the accurate alignment of material properties over large areas. Meter-sized ORISs can lead to substantial delay spreads, particularly when there is a large discrepancy between the incident and reflected angles \cite{Trinh2024-JSAC}. As a result, careful consideration of both the ORIS size and system symbol duration is crucial to mitigate delay spread-induced inter-symbol interference.
\section{The Road Ahead}
\subsection{Transmissive ORIS Facilitating Full-Space Coverage}
The transmissive ORIS, leveraging translucent liquid crystal cells, represents a breakthrough in improving the angular quantum coverage. This technology allows optical beams to pass through the reconfigurable surface, while accurately manipulating their phase, amplitude, and polarization, providing sophisticated control over both the beam diameter and power distribution. When positioned vertically along the edges of building rooftops, the transmissive ORIS enables a SAGIN platform to direct optical beams through its surface, beneficially deflecting them either upward or downward upon exit. This capability allows for the beam to be directed towards targets below the height of the ORIS, effectively improving the spatial coverage provided by reflective ORIS systems, which typically reflect the beams upward, as illustrated in Fig. \ref{fig_1}.

Moreover, high-order optical beams or high-dimensional quantum photons can be generated using a transmissive ORIS by applying a hologram or phase mask specifically designed for imposing precise phase shifts or amplitude variations across the wavefront of the incident light as it passes through the surface. This process encodes complex spatial modes or higher-dimensional quantum states into the beam. The transmissive ORIS, which can be scaled to match the size of the incident beam along the propagation path, or miniaturized to be integrated into optical quantum terminals, facilitates the development of reconfigurable intelligent transceivers for advanced NG communication systems. Despite its considerable potential, the application of transmissive ORISs in both classical and quantum communications remains largely unexplored, highlighting a promising new avenue and a critical component for future ORIS-aided SAGIN systems. 
\subsection{ORIS-Assisted Power Transfer for UAV Recharging}
The integration of ORIS in wireless power transfer systems presents a promising solution for recharging battery-powered UAVs within NG SAGIN environments. ORIS-assisted power transfer allows for precise and dynamic control of high-power optical beams, specifically designed for efficient wireless energy transfer to UAVs. By skillfully steering and focusing these energy beams, ORIS maximizes power delivery efficiency, ensuring that UAVs remain charged and fully operational during extended missions. This capability is particularly critical in SAGIN scenarios, where UAVs function as essential links among space, air, and ground nodes. The ability to wirelessly recharge UAVs mid-flight not only extends their operational range but also eliminates the need for frequent landings, ensuring continuous, uninterrupted support for both classical and quantum communication networks. Additionally, this approach offers the significant advantage of directing energy beams upward, perfectly aligned with the high-altitude flight paths of UAVs, while ensuring the safety of people on the ground.

However, several challenges must be addressed to realize the full potential of ORIS-assisted power transfer for UAV recharging. A significant challenge is the accurate alignment and real-time tracking of UAVs to ensure that the energy beams are efficiently directed, especially in dynamic environments with moving targets. This requires precise control systems and advanced algorithms capable of responding to changes in UAV position and orientation. Additionally, maintaining the integrity of quantum communication channels during power transfer is another challenge, as the process may impose noise or decoherence that could perturb the delicate quantum states. There is also the issue of energy loss over long distances, particularly in hostile atmospheric conditions, which could reduce the efficiency of power transfer. Addressing these challenges will require substantial research and development, including the optimization of ORIS design, the integration of robust tracking systems, and the development of protocols to mitigate potential disruptions in quantum communications.
\subsection{Software-Defined Networks with Artificial Intelligence}
The integration of software-defined networks (SDNs) with artificial intelligence (AI) is set to revolutionize the practical operation of NG SAGINs. SDNs offer flexible centralized control of network resources \cite{Cao2022}, enabling real-time adjustments that are critical for managing the dynamic and heterogeneous environments of SAGINs. When combined with AI, these networks become capable of autonomously optimizing resource allocation, traffic management, and security, ensuring that the network remains responsive even as it evolves with the addition, grouping, and movement of satellites and UAVs. This adaptability is particularly crucial for supporting the stringent demands of quantum communications, where both maintaining signal coherence and minimizing latency are essential. By leveraging AI, SDNs can predict network demands, preemptively address bottlenecks, and implement proactive strategies, thereby ensuring seamless communications across the complex, multi-layered architecture of NG SAGIN. 

The integration of AI into SDN management presents several significant challenges. One of the foremost issues is the complexity of AI models needed for managing the diverse rapidly evolving conditions within SAGIN environments. Developing algorithms that can accurately adapt to these conditions in a near-real-time fashion is a daunting task, as instability in SAGIN platforms can perturb entangled states or cause decoherence. This requires advanced machine learning techniques and substantial computational power for effectively monitoring, predicting, and mitigating these perturbations. Additionally, the global deployment of AI-driven SDNs raises concerns about scalability, as the networks must maintain efficiency across vast and varied geographical regions, while minimizing the potential escalation of latency. Security is also a critical concern. Albeit AI is capable of bolstering network defenses, it can also be exploited by adversaries to launch sophisticated attacks. Furthermore, the continuous operation of AI algorithms across large-scale networks like SAGINs demands significant energy resources, potentially negating the efficiency gains achieved through intelligent network management. Overcoming these challenges will necessitate future research and innovation to fully harness the potential of AI-enhanced SDNs in NG SAGINs.
\section{Conclusions}
SAGIN is set to transform NG networks by seamlessly integrating satellites, UAVs, manned aircraft, and the terrestrial infrastructure to deliver resilient, ubiquitous, and global connectivity. By incorporating quantum communications through optical wireless signals, a global Qinternet can be established alongside traditional networks. However, challenges such as LoS requirements and beam broadening may impede long-distance optical beam propagations. We highlighted that deploying ORISs on building rooftops offers a solution by enabling adaptive beam control and blockage-mitigation. Our initial findings underscored the crucial role that ORISs will play in optimizing quantum performance. Future directions include deploying transmissive ORISs alongside their reflective counterparts to enhance spatial coverage and utilizing ORISs for wireless power transfer to battery-powered UAVs. Additionally, we discussed the application of AI-driven SDNs for the autonomous management and optimization of the upcoming NG quantum SAGINs.


\section*{Biographies}
\vspace{-33pt}
\begin{IEEEbiographynophoto}{Phuc V. Trinh} (Senior Member, IEEE) (trinh@iis.u-tokyo.ac.jp) received the B.E. degree in electronics and telecommunications from the Posts and Telecommunications Institute of Technology, Hanoi, Vietnam, in 2013, and the M.Sc. and Ph.D. degrees in computer science and engineering from The University of Aizu, Aizuwakamatsu, Japan, in 2015 and 2017, respectively. From 2017 to 2023, he was a Researcher with the Space Communication Systems Laboratory, National Institute of Information and Communications Technology, Tokyo, Japan. Since 2023, he has been a Project Research Associate with the Institute of Industrial Science, The University of Tokyo, Tokyo, Japan. His current research interests include optical and wireless communications for space, airborne, and terrestrial networks.
\end{IEEEbiographynophoto}

\begin{IEEEbiographynophoto}{Shinya Sugiura} (Senior Member, IEEE)  (sugiura@iis.u-tokyo.ac.jp) received the B.S. and M.S. degrees in aeronautics and astronautics from Kyoto University, Japan, in 2002 and 2004, respectively, and the Ph.D. degree in electronics and electrical engineering from the University of Southampton, U.K., in 2010. He was a Research Scientist with Toyota Central R\&D Laboratories, Inc., Japan, from 2004 to 2012 and an Associate Professor with Tokyo University of Agriculture and Technology, Japan, from 2013 to 2018. Since 2018, he has been with the Institute of Industrial Science, The University of Tokyo, Japan, where he is currently a full Professor. His research has covered wireless communications and signal processing.
\end{IEEEbiographynophoto}


\begin{IEEEbiographynophoto}{Chao Xu} (Senior Member, IEEE) (cx1g08@ecs.soton.ac.uk) received the B.Eng. degree in telecommunications from the Beijing University of Posts and Telecommunications, China, the B.Sc. (Eng.) degree (Hons.) in telecommunications from the Queen Mary University of London, U.K., through the Sino-U.K. Joint Degree Program, in 2008, and the M.Sc. degree (Hons.) in radio frequency communication systems and the Ph.D. degree in wireless communications from the University of Southampton, U.K., in 2009 and 2015, respectively. He is currently a Senior Lecturer with the Next Generation Wireless Research Group, University of Southampton. His research interests include index modulation, reconfigurable intelligent surfaces, noncoherent detection, and turbo detection. 
\end{IEEEbiographynophoto}

\begin{IEEEbiographynophoto}{Lajos Hanzo} (Life Fellow, IEEE) (lh@ecs.soton.ac.uk) received his degree in electronics in 1976 and his doctorate in 1983. He holds an honorary doctorate from the Technical University of Budapest (2009) and from the University of Edinburgh (2015). He is a member of the Hungarian Academy of Sciences and a former Editor-in-Chief of the IEEE Press. 
\end{IEEEbiographynophoto}

\vfill


\begin{thebibliography}{}
\providecommand{\url}[1]{#1}
\csname url@samestyle\endcsname
\providecommand{\newblock}{\relax}
\providecommand{\bibinfo}[2]{#2}
\providecommand{\BIBentrySTDinterwordspacing}{\spaceskip=0pt\relax}
\providecommand{\BIBentryALTinterwordstretchfactor}{4}
\providecommand{\BIBentryALTinterwordspacing}{\spaceskip=\fontdimen2\font plus
\BIBentryALTinterwordstretchfactor\fontdimen3\font minus
  \fontdimen4\font\relax}
\providecommand{\BIBforeignlanguage}[2]{{%
\expandafter\ifx\csname l@#1\endcsname\relax
\typeout{** WARNING: IEEEtran.bst: No hyphenation pattern has been}%
\typeout{** loaded for the language `#1'. Using the pattern for}%
\typeout{** the default language instead.}%
\else
\language=\csname l@#1\endcsname
\fi
#2}}
\providecommand{\BIBdecl}{\relax}
\BIBdecl

\end{thebibliography}


\begin{thebibliography}{99}
\bibliographystyle{IEEEtran}
\bibitem{Wang2024}
T. Wang et~al.,
``Quantum-empowered federated learning in space-air-ground integrated networks," 
\emph{IEEE Netw.}, vol. 38, no. 1, pp. 96--103, Jan. 2024.

\bibitem{Xu2019}
C. Xu et~al.,
``Adaptive coherent/non-coherent spatial modulation aided unmanned aircraft systems," 
\emph{IEEE Wirel. Commun.}, vol. 26, no. 4, pp. 170--177, Aug. 2019.

\bibitem{Chen2024}
L. Chen, K. Xue, J. Li, Z. Li, and N. Yu,
``Q-CSKDF: A continuous and security key derivation function for quantum key distribution," 
\emph{IEEE Netw.}, vol. 38, no. 5, pp. 123--130, Sept. 2024.

\bibitem{Trinh2024}
P. V. Trinh and S. Sugiura, 
``Quantum Internet in the sky: vision, challenges, solutions, and future directions," 
\emph{IEEE Commun. Mag.}, vol. 62, no. 10, pp. 62--68, Oct. 2024.

\bibitem{Li2021}
Z. Li et~al.,
``Building a large-scale and wide-area quantum Internet based on an OSI-alike model," 
\emph{China Commun.}, vol. 18, no. 10, pp. 1--14, Oct. 2021.

\bibitem{Li2023}
Z. Li et~al.,
``Entanglement-assisted quantum networks: Mechanics, enabling technologies, challenges, and research directions," 
\emph{IEEE Commun. Surveys Tuts.}, vol. 25, no. 4, pp. 2133--2189, 4th Quart., 2023.

\bibitem{Ndjiongue2021}
A. R. Ndjiongue, T. M. N. Ngatched, O. A. Dobre and H. Haas,
``Design of a power amplifying-RIS for free-space optical communication systems," 
\emph{IEEE Wirel. Commun.}, vol. 28, no. 6, pp. 152--159, Dec. 2021.

\bibitem{Ajam2024}
H. Ajam, M. Najafi, V. Jamali, and R. Schober,
``Optical IRSs: power scaling law, optimal deployment, and comparison with relays," 
\emph{IEEE Trans. Commun.}, vol. 72, no. 2, pp. 954--970, Feb. 2024.

\bibitem{Kundu2024}
N. K. Kundu, M. R. McKay, R. Murch, and R. K. Mallik,
``Intelligent reflecting surface-assisted free space optical quantum communications," 
\emph{IEEE Trans. Wirel. Commun.}, vol. 23, no. 5, pp. 5079--5093, May 2024.

\bibitem{Trinh2024-JSAC}
P. V. Trinh, S. Sugiura, C. Xu, and L. Hanzo,
``Optical RISs improve the secret key rate of free-space QKD in HAP-to-UAV scenarios," 
\emph{IEEE J. Sel. Areas Commun.}, to be published.

\bibitem{Azuma2023}
 K. Azuma et~al.,
``Quantum repeaters: From quantum networks to the quantum Internet," 
\emph{Rev. Mod. Phys.}, vol. 95, no. 4, Dec. 2023, Art. no. 045006.

\bibitem{Liu2021}
H.-Y. Liu et~al.,
``Optical-relayed entanglement distribution using drones as mobile nodes," 
\emph{Phys. Rev. Lett.}, vol. 126, no. 2, Jan. 2021, Art. no. 020503.

\bibitem{Khatri2021}
S. Khatri et~al.,
``Spooky action at a global distance: analysis of space-based entanglement distribution for the quantum Internet," 
\emph{npj Quantum Inf.}, vol. 7, no. 1, Jan. 2021, Art. no. 4.

\bibitem{Pirandola2021}
S. Pirandola, R. Laurenza, C. Ottaviani, and L. Banchi, 
``Fundamental limits of repeaterless quantum communications," 
\emph{Nature Commun.}, vol. 8, no. 1, Apr. 2017, Art. no. 15043.

\bibitem{Cao2022}
Y. Cao, Y. Zhao, J. Zhang and Q. Wang,
``Software-defined heterogeneous quantum key distribution chaining: an enabler for multi-protocol quantum networks," 
\emph{IEEE Commun. Mag.}, vol. 60, no. 9, pp. 38--44, Sept. 2022.

\end{thebibliography}
\end{document}